\documentstyle[12pt]{article}
\begin{document}
\setlength{\unitlength}{1mm}

{\hfill   August 1997 }

{\hfill    WATPHYS-TH-97/13}

{\hfill    hep-th/9709064} \vspace*{2cm} \\
\begin{center}
{\Large\bf Universality of Quantum Entropy for Extreme Black Holes}
\end{center}
\begin{center}
{\large\bf Robert B.~Mann\footnote{e-mail: mann@avatar.uwaterloo.ca} 
and 
Sergey N.~Solodukhin\footnote{e-mail: sergey@avatar.uwaterloo.ca}}
\end{center}
\begin{center}
{\it Department of Physics, University of Waterloo, Waterloo, Ontario 
N2L 3G1, Canada}  
\end{center}
\vspace*{1cm}
\begin{abstract}
\noindent
We consider the extremal limit of a black hole geometry
of the Reissner-Nordstrom  type and compute the quantum
corrections to its entropy.
Universally, the limiting geometry is the direct product of two 2-dimensional
 spaces and is characterized by just a few parameters. We argue
that the quantum corrections to the entropy of such extremal black holes
due to a massless scalar field have a universal behavior.
We obtain explicitly the form of the quantum entropy in this extremal limit 
as function of the parameters of the limiting geometry. We generalize
these results to black holes with toroidal or higher genus horizon topologies.
In general, the extreme quantum entropy 
is completely determined by the spectral geometry of the horizon
and in the ultra-extreme case it is just a determinant of the 2-dimensional
Laplacian. As a byproduct of our considerations we obtain expressions for the
quantum entropy of black holes which are not of the Reissner-Nordstrom type: the
extreme dilaton and extreme Kerr-Newman black holes. In both cases the classical
Bekenstein-Hawking entropy is modified by logarithmic corrections. 
\end{abstract}
\newpage
\baselineskip=.8cm

The origin of black hole entropy remains an exciting unresolved
problem requiring implementation
of new ideas and deep insights. Of particular interest is the physics of 
extremal black holes and their entropy. Taking the extremal limit directly 
in the Bekenstein-Hawking (BH) 
expression \cite{1} on finds that entropy of an extreme black hole is
proportional to the area of its horizon 
\begin{equation}
S_{BH}^{ext}={A\over 4G}
\label{1}
\end{equation}
as it is in the non-extreme case. The same situation holds for the quantum
corrections to the extremal BH entropy. In particular, quantum corrections due
to a minimal scalar field yield UV-divergent contributions to the entropy of a
Reissner-Nordstrem (RN) black hole which in the extreme limit 
take the form \cite{2}
\begin{equation}
S^{ext}_q={A\over 48\pi\epsilon^2}+{1\over 90}\ln\left(\frac{\epsilon}{\Lambda}\right)~~,
\label{2}
\end{equation}
where $\epsilon$ is the ultraviolet (UV) cut-off and $\Lambda$ is some
length scale associated with the black hole metric.

However, it was proposed recently \cite{3}
that extreme black holes actually must have zero entropy, $S^{ext}=0$, 
and can be in thermal equilibrium at an arbitrary temperature.
The reason for this is purely topological. In the extreme case the
Euclidean space-time near the horizon has the form of an infinite throat that
allows one to identify the Euclidean time variable with an arbitrary 
period $T^{-1}$, thereby implying that its Lorentzian continuation corresponds 
to an extremal black hole with an arbitrary temperature $T$.  This differs
from the non-extreme case where for Euclidean time with an arbitrary period
$T^{-1}$, a conical singularity at the horizon is present in the Euclidean section,
disappearing only for the Hawking temperature $T=T_H$.

However subsequent study of extremal black holes in superstring theory \cite{4}
indicates that their entropy can be understood to arise from degrees of freedom
that correspond to basic string states. This gives strong
support to the first viewpoint, namely that the entropy of an extreme black hole is
proportional to its horizon area.  Whether or not this viewpoint is consistent with
the topological arguments of ref. \cite{3} remains unclear.

Recently,  Zaslavskii \cite{5} proposed a method for reaching the extremal
limit of a black holes whilst remaining in the topological
class of the corresponding non-extremal geometries. He considered a sequence of
non-extreme black holes in a cavity at $r=r_B$ and found that there 
exists a set of data $(r_+,r_B,r_-)$ such that the limit
${r_+\over r_-}\rightarrow 1,~{r_B\over r_+}\rightarrow 1$
is well-defined. The resultant limiting geometry belongs to
the non-extremal topological class and takes the form of a direct product 
of two 2-dimensional spaces. Moreover, there is  
a remarkable universality. No matter how arbitrary the non-extreme 
geometry is, the limiting extremal one
is characterized by just a few parameters: a pair of  dimensional 
ones, $r_+$ and $\bar{b}$,  and one dimensionless parameter $c$. 
For $c=1$ the limiting geometry is the direct product of 2-dimensional hyperbolic 
space $(H_2)$ with the two-dimensional sphere $(S_2)$, where
($\bar{b}^{-1/2}$ and $r_+$ are the respective radii of $H_2$ and $S_2$).
Since the limiting geometry belongs
to the non-extreme class its entropy is necessarily
proportional to the horizon area in accord with the BH formula.

The classical BH entropy (\ref{1}) is actually a tree-level quantity.
In a theory of gravity whose Lagrangian contains terms quadratic in curvature, 
the entropy is modified by terms which are not proportional to the horizon area
and depend upon the external geometry of the horizon \cite{55/2}.
Any quantum field propagating in the black hole background adds
some (one-loop) correction $S_q$ to the entropy. Generally, this correction
is not determined only by the horizon geometry and is 
a complicated functional of the background metric. Very little is known about 
it in general, and the exact form of
this functional can be retrieved only in some special cases.
In particular, it is known in two dimensions \cite{51/2} for 
conformal scalar matter and in three dimensions \cite{6} for a scalar field 
on the BTZ background.

The universality we mentioned above suggests that
the quantum entropy $S^{BH}$of a black hole possesses a
universal behavior in its extreme limit going to some limiting 
function $S^{BH}\to S^{ext}_q(r_+,b,c)$.
The fact that for $c=1$ the limiting geometry has an especially simple form,
$H_2 \times S_2$, suggests that in this case the limiting function
$S^{ext}_q(r_+,b)$ can be found explicitly. 

The purpose of this paper is to demonstrate these results.
We  start first with a brief review of the limiting procedure of \cite{5} and then 
calculate the effective action and the quantum entropy due to a minimal
scalar field. We use the heat kernel technique which for the present problem
proceeds in a manner similar to that developed in ref. \cite{6} for the $(2+1)$-dimensional 
BTZ black hole. We refer the reader to the ref. \cite{6} for technical details.

Before proceeding, we pause to note that
an earlier attempt \cite{14b} at computing quantum corrections to the entropy of 
topologically extremal 
Reissner-Nordstrom black holes \cite{3} found that these were proportional to
powers of the temperature and had divergences that were no stronger than
in the non-extremal case.  The renormalization approach employed
implied that the temperature necessarily vanished, suggesting
that these quantum corrections to the entropy in the extremal case also vanished.
However such a value is not a smooth continuation of the non-extremal case.

Consider an arbitrary static spheri-symmetric  geometry
\begin{eqnarray}
&&ds^2=g(r)d\tau^2+{1\over f(r)}dr^2+r^2d\omega^2~~, \nonumber \\
&&d\omega^2=d\theta^2+\sin^2\theta d\varphi^2~~,
\label{3}
\end{eqnarray}
describing a non-extreme hole with an outer horizon located at $r=r_+$. Then,
the functions $g(r)$ and $f(r)$ in (\ref{3}) can be expanded as follows
\begin{eqnarray}
&&f(r)=a(r-r_+)+b(r-r_+)^2+O((r-r_+)^2)~~ \nonumber \\
&&g(r)=g_0(r-r_+)+g_1(r-r_+)^2+O((r-r_+)^2)~~.
\label{4}
\end{eqnarray}
It is convenient to consider the geodesic distance $l=\int f^{-1/2}dr$ as a
radial coordinate. Retaining the first two terms in the first of
equations (\ref{4}), we find for $r>r_+$ that
\begin{equation}
(r-r_+)={a\over b} \sinh^2 ({lb^{1/2}\over 2})
\label{5}
\end{equation}
and that
\begin{equation}
g(lb^{1/2})=({a\over b})^2g_1\sinh^2({lb^{1/2}\over 2})
(c+\sinh^2({lb^{1/2}\over 2}))~~,
\label{6}
\end{equation}
where $c={g_0b/ a g_1}$ is a dimensionless parameter.
In order to avoid appearance of a conical singularity at $r=r_+$
the Euclidean time $\tau$ in (\ref{3}) must be identified with
period $4\pi/\sqrt{ag_0}$
which goes to infinity in the extreme limit
$a\rightarrow 0$. However, rescaling $\tau \rightarrow \phi=\tau{\sqrt{ag_0}/ 2}$
yields a new variable $\phi$ having period $2\pi$. Then, taking into
account (\ref{5}) and (\ref{6}) we find for the metric (\ref{3})
\begin{eqnarray}
&&ds^2=\frac{1}{b}\left(g(x)d\phi^2+dx^2\right)
+(r_++{a\over b}\sinh^2{x\over 2})^2d\omega^2~~,
\nonumber \\
&&g(x)=4\sinh^2({x\over 2})(1+\frac{1}{c}\sinh^2{x\over 2})~~,
\label{7}
\end{eqnarray}
where we have introduced the variable $x=lb^{1/2}$.

To obtain the extremal limit we take $a\rightarrow 0$ keeping
$c$ finite (i.e. the ratio $a/ g_0$ must remain finite).
The limiting geometry 
\begin{equation}
ds^2=\frac{1}{b}\left(g(x)d\phi^2+dx^2\right)+ r^2_+d\omega^2
\label{8}
\end{equation}
is that of the direct product of two 2-dimensional spaces and is
characterized by a pair of dimensional parameters
$b^{-1/2}$ and $r_+$ and one dimensionless parameter $c$.
The parameter $r_+$ sets the length scale for the 2-dimensional
space of constant curvature described by the 2-metric $d\omega^2$. 
The parameter $b^{-1/2}$
performs the same job for the $(x,\phi)$ 2-space, and the
parameter $c$ provides a dimensionless measure of how far this
2-space departs from one of constant negative curvature.
This is the universality we mentioned above: although the non-extreme geometry
is in general described by an infinite number of parameters associated with the
determining functions $f(r)$ and $g(r)$ the geometry in the extreme limit 
depends only on three parameters ${b},~r_+,~c$. 
Note that the coordinate $r$ is inadequate for describing the
extremal limit (\ref{8}) since the coordinate transformation (\ref{5})
is singular when $a\rightarrow 0$.

The limiting geometry (\ref{8}) takes especially simple form if 
$g(r)=f(r)$. Then $c=1$ and the extreme metric
\begin{equation}
ds^2={1\over b}(\sinh^2 xd\phi^2+dx^2)+r^2_+d\omega^2
\label{9}
\end{equation}
is that of a direct product $H_2\times S_2$ of 2d hyperbolic space $H_2$ with
radius $l=b^{-1/2}$ and a 2d sphere $S_2$ with radius $l_1=r_+$.
We shall consider this case below. It is worth noting that the limiting geometry
(\ref{8}) (or (\ref{9})) precisely merges near the horizon with the
geometry of the original metric (\ref{3}) in the sense that all the 
curvature tensors for both metrics coincide.
This is in contrast with, say, the situation in which the Rindler metric is considered
to approximate to geometry of a non-extreme black hole:
the curvatures of both spaces do not merge in general.

For the Reissner-Nordstrom black hole we have $b^{-1}=r^2_+$ and the limiting 
extreme geometry is the well-known 
Bertotti-Robinson space characterized by just one parameter $r_+$.
This space has remarkable properties in the context of supergravity theory. 
In particular, the effective action of $N=2$ supergravity was
shown \cite{63/2} to acquire no quantum corrections in the background
of this metric. These supersymmetry aspects, though intriguing,
are not the subject of interest in this paper and we shall not consider them 
any further.

In the past year there has been an active study of generalizations of the metric 
(\ref{3}) which describe black holes with horizons of non-trivial topology \cite{61/2}.
In this generalization the spherical element 
$d\omega^2$ in (\ref{3}) is replaced by either the flat element
\begin{equation}
d\omega^2_T=d\theta^2+d\varphi^2
\label{x}
\end{equation}
or the hyperbolic element 
\begin{equation}
d\omega^2_H=d\theta^2+\sinh^2\theta d\varphi^2
\label{xx}
\end{equation}
describing a space with constant negative curvature, and the coordinates
in this space are appropriately identified so as to make the space both
compact and free of conical singularities.  Remarkably, the metric 
(\ref{3}) with such ``angle'' elements
is still a solution of the Einstein equations with a cosmological
term. Asymptotically, however, it does not have the topology of
flat space. After the appropriate identifications in the $(\theta , \varphi )$
section, the metrics (\ref{x}) and (\ref{xx}) describe respectively a 
torus $T=\Sigma_{g=1}$ and a surface $\Sigma_g$ of genus $g\geq 2$ 
respectively \cite{61/2}.
Since the element (\ref{x}) or (\ref{xx}) describes the metric on the horizon
we observe that the horizon may have a non-trivial topology characterized
by an arbitrary genus $g$. The limiting procedure described above goes through
for these types of black holes and in general
we obtain the limiting metric
\begin{equation}
ds^2={1\over b}(g(x)d\phi^2+dx^2)+r_+^2d\omega^2_{(g)}~~,
\label{xxx}
\end{equation}
where $d\omega^2_{(g)}$ is the metric on a 2d compact surface of genus $g$.
The extreme geometry now is $E_g=H_2\times\Sigma_g$.
All our formulas  may be generalized to include arbitrary $g$.

Note also that not all the known black holes are described by a metric of the 
form (\ref{3}). Exceptions include static dilaton black holes and the
stationary Kerr-Newman black holes. For these cases the limiting procedure
must be reformulated. However, insofar as the entropy 
calculation is concerned, our computations below provide us with a hint 
as to how in some cases the total renormalized entropy can be obtained even 
without knowledge of the limiting geometry. One needs only to know the UV
divergences of the quantum entropy. These have been obtained  
in the earlier studies \cite{2}, \cite{10}, \cite{14a}, \cite{14b},
\cite{14c}.

\bigskip

Consider now a scalar field minimally coupled to gravity.
Quantizing it on the black hole background (\ref{9}) yields 
quantum corrections to the black hole entropy. It is these corrections
that we wish to compute for extremal black holes.
We proceed by allowing the coordinate $\phi$ which plays the
role of Euclidean time to have period 
$2\pi\alpha$. For $\alpha\neq 1$ the metric (\ref{9}) then describes the
space $E^\alpha = H^\alpha_2\times S_2$, where $H^\alpha_2$
is the hyperbolic space coinciding with $H_2$ everywhere except the 
point $x=0$ where it has a conical singularity with an
angular deficit $\delta=2\pi (1-\alpha )$.

The effective action of the scalar field 
\begin{equation}
W_{eff}=-{1\over 2}\int^\infty_{\epsilon^2}{ds\over s} Tr K
\label{10}
\end{equation}
is expressed by means of the 
trace of the heat kernel $K(z,z',s)$ satisfying the heat
equation
\begin{eqnarray}
&&(\partial_s-\Box ) K(z,z',s)=0~~, \nonumber \\
&&K(z,z',s=0)=\delta(z,z')~~,
\label{11}
\end{eqnarray}
where $\Box=\nabla_\mu\nabla^\mu$ is the Laplace operator.
A nice property of the heat kernel is that it  factorizes
on a product space. Thus, for the heat kernel on $E^\alpha$
we have
$$
K_{E^\alpha}(z,z',s)=K_{H_2^\alpha}(x,x',\phi , \phi ',s)~K_{S^2}(\theta ,
\theta ', \varphi, \varphi ',s)~~
$$
and the effective action reads
\begin{equation}
W_{eff}[E^\alpha ]=-{1\over 2}\int^\infty_{\epsilon^2}{ds\over s} Tr K_{H_2^\alpha} Tr K_{S_2}~~,
\label{12}
\end{equation}
where $\epsilon$ is a UV cut-off.
On spaces with constant curvature the heat kernel function 
is known explicitly \cite{7}. In particular, on a
2d space with constant negative curvature $R=-{2\over l^2}$:
\begin{equation}
K_{H_2}(z,z',s)={1\over l^2}{\sqrt{2}e^{-{\bar{s}/ 4}}
\over (4\pi \bar{s})^{3/2}}\int_\sigma^\infty {dyye^{-{y^2/ 4\bar{s}}}\over
\sqrt{\cosh y-\cosh \sigma }}~~,
\label{13}
\end{equation}
where $\bar{s}=sl^{-2}$.
The corresponding expression on a 2d sphere (which has
constant positive curvature $R={2\over l^2_1}$) is
\begin{equation}
K_{S_2}(z,z',s)={1\over l^2_1}{\sqrt{2}e^{-{\bar{s}/ 4}}
\over (4\pi \bar{s})^{3/2}}\sum_{n=-\infty}^\infty (-1)^n
\int^\pi_\sigma {d\varphi (\varphi+2\pi n)e^{-{(\varphi+2\pi n)^2\over 4 \bar{s}}} \over \sqrt{\cos \sigma-\cos \varphi }}~~,
\label{14}
\end{equation}
where now $\bar{s}=sl^{-2}_1$.
Alternatively it can be represented in terms of the Legendre polynomials
$P_n(\cos \sigma )$ as follows
\begin{equation}
K_{S_2}(z,z',s)={1\over 4\pi l^2_1} \sum_{n=0}^\infty (2n+1) P_n(\cos \sigma )
e^{-{\bar{s}n(n-1)}}~~.
\label{15}
\end{equation}
The trace of the heat kernel on  2d sphere then takes the form
\begin{eqnarray}
&&TrK_{S_2}(s)=\Theta_S (\bar{s}{l^2\over l_1^2}) ~~,\nonumber \\
&&\Theta_S (s) \equiv \sum_{n=0}^\infty
(2n+1) e^{-sn(n+1)}~~.
\label{20}
\end{eqnarray}

In eqs.(\ref{13}) and (\ref{14}), (\ref{15}) $\sigma$ is the geodesic
distance between  
the points $z$ and $z'$ on $H_2$ and $S_2$ respectively. In particular, on $H_2$ the
geodesic distance $\sigma$ between the points $(x, \phi )$ 
and $(x, \phi+\Delta \phi )$ is given by
$$
\sinh^2 {\sigma \over 2}=\sinh^2 x \sin^2 {\Delta \phi \over 2}~~.
$$

In principle, there must be a boundary at $x=x_B$ where some boundary 
conditions are imposed on the quantum scalar field.
This would modify the heat kernel function (\ref{13}). The boundary, however,
can be arbitrarily far from the horizon at $x=0$.
Since the entropy of the hole itself comes exclusively from the conical 
singularity at $x=0$ the boundary at $x=x_B$ will not to affect the entropy
calculation. Therefore, we neglect the boundary and deal with the 
infinite space $H_2$.

Hence the problem of evaluating the effective action (\ref{12}) on $E^\alpha$
reduces to that of evaluating the heat kernel on the conical space 
$H_2^\alpha$. To find it we employ the Sommerfeld formula \cite{8}
\begin{equation}
K_{H^\alpha_2}(z,z',s)=K_{H_2}(z,z',s)+{\imath \over 4\pi\alpha}
\int_\Gamma \cot {w\over 2\alpha}K_{H_2}(\phi-\phi '+w,s)dw~~,
\label{16}
\end{equation}
where $K_{H^2}$ is the heat kernel on  regular manifold.
The contour $\Gamma$ in 
(\ref{16}) consists of two vertical
lines, going from $(-\pi+\imath \infty )$ to $(-\pi-\imath \infty )$
and from $(\pi-\imath \infty )$ to $(\pi+\imath \infty )$ and
intersecting the real axis between the poles of the $\cot {w\over 2\alpha}$:
$-2\pi\alpha,~0$ and $0,~+2\pi\alpha$ respectively. 
For $\alpha=1$ the integrand in (\ref{16}) is 
a $2\pi$-periodic function and the contributions from these two vertical 
lines (at a fixed distance $2\pi$ along the real axis) 
cancel each other.
 Applying the formula (\ref{16}) we first define the quantity
$$
Tr_w K_{H_2}=\int_{H_2}K_{H_2}(x=x', \phi'=\phi+w)d\mu_x~~,
$$
where $d\mu_x=l^2\sinh{x} dxd\phi$ is the measure on $H_2$. Inserting 
(\ref{13}) into this expression, interchanging the $y$ and $x$
limits of integration, performing the $x$ and $\phi$
integrations,  and then integrating by parts on $y$ yields
\begin{equation}
Tr_w K_{H_2}={e^{-{\bar{s}/4}}
\over (4\pi \bar{s})^{1/2}}\int_0^\infty {dy \cosh y e^{-{y^2/ \bar{s}}}
\over (\sinh^2y+\sin^2{w\over 2})}~~.
\label{17}
\end{equation}
Then the trace of  the heat kernel on $H^\alpha_2$ can be evaluated as
$$
Tr K_{H^\alpha_2}=\alpha Tr K_{H_2}+{\imath \over 4\pi}
\int_\Gamma dw\cot {w\over 2\alpha}~Tr_w K_{H_2}~~.
$$
The contour integration is easily performed using the formula
\begin{equation}
{\imath \over 4\pi\alpha}\int_\Gamma {\cot{w\over 2\alpha}dw
\over \sinh^2y+\sin^2{w\over 2}}=f(y,\alpha )\equiv
{1\over \sinh^2y}({1\over \alpha} {\tanh y\over \tanh {y\over \alpha}}-1)~~,
\label{18}
\end{equation}
obtained in \cite{6}.
Then we arrive at an integral representation for
the heat kernel on $H^\alpha_2$:
\begin{equation}
TrK_{H^\alpha_2}=\alpha TrK_{H_2}+\alpha {e^{-{\bar{s}/ 4}}
\over (4\pi \bar{s})^{1/2}}\int_0^\infty dy \cosh y f(y,\alpha ) 
e^{-{y^2/ \bar{s}}}~~,
\label{19}
\end{equation}
where $\bar{s}=s l^{-2}$. We are mainly interested in the second
term in (\ref{19}) which makes a contribution to the entropy.
Using  (\ref{19}) and (\ref{20}) we may find an integral
representation for the effective action $W_{eff}[E^\alpha ]$ (\ref{12}).
The quantum entropy $S_q$ can be calculated from $W_{eff}[E^\alpha ]$
by making use the known formula
\begin{equation}
S_q=(\alpha \partial_\alpha-1)W_{eff}[E^\alpha ]|_{\alpha=1}~~.
\label{en}
\end{equation}
Applying it to our case it is useful to note that the function
(\ref{18}) for $\alpha \simeq 1$ behaves as
$$
f(y,\alpha )= {1\over \sinh^2y}(1-{2y\over \sinh 2y}) (1-\alpha )
+O((1-\alpha )^2)~~.
$$
Then the quantum entropy is easily found to have   the form
\begin{equation}
S_q^{ext}={1\over 4\sqrt{\pi}}\int^\infty_{\epsilon^2 / l^2} {ds\over s^{3/2}} 
k_H (s)\Theta_S (s {l^2\over l_1^2})e^{-{s/ 4}}~~,
\label{21}
\end{equation}
where
\begin{equation}
 k_H (s)=\int^\infty_0 dy {\cosh y\over \sinh^2y}(1-{2y\over \sinh 2y})
e^{-{y^2/ s}}~~.
\label{22}
\end{equation}
The decrease of the integrand in (\ref{21}) for large $s$ so rapid there
is no need to introduce an infra-red cut-off to make the integral 
(\ref{21}) convergent at large $s$.
We see that the entropy $S_q$ (\ref{21}) is a positive quantity which
contains a divergent contribution in the limit $\epsilon \rightarrow 0$. 
This divergence can be isolated by using the asymptotic expansions
$$
k_H(s)= \sqrt{\pi s}({1\over 3}-{s\over 20}+O(s^2))~~,
$$
and
$$
\Theta_S (s)= {1\over s} (1+{s\over 3}+O(s^2))
$$
yielding
\begin{equation}
S^{ext}_{div}={1\over 12} {l_1^2\over \epsilon^2} +({1\over 18}-{1\over 15}
{l_1^2\over l^2})\ln {l\over \epsilon }~~
\label{23}
\end{equation}
for the UV divergent contribution to (\ref{21}).

Recall that in terms of the limiting metric (\ref{9}) we have
$l_1=r_+$ and $l=b^{-1/2}$. For an extremal Reissner-Nordstrom black hole
$l=l_1=r_+$ and this expression becomes
\begin{equation}
S^{ext}_{div}={1\over 12} {r_+^2\over \epsilon^2}-{1\over 90}\ln {r_+\over\epsilon}
\label{24'}
\end{equation}
and coincides with the result (\ref{2}) found earlier in \cite{2}.

According to general scheme described in ref. \cite{9}, the divergence 
(\ref{23}) can be absorbed into the renormalization of the coupling constants 
in the gravitational effective action. After this procedure the total entropy 
$S=S_{BH}+S_q$ takes the form
\begin{equation}
S^{ext} (l,l_1)={A_+\over 4G}+({1\over 18}-{1\over 15}
{l_1^2\over l^2})\ln {l\over \mu}+\Xi ({l_1\over l})~~,
\label{24}
\end{equation}
where $\mu$ is a dimensional parameter, 
$A_+=4\pi l_1^2$ is area of the horizon and the last term depends only 
on the ratio $\kappa={l^2_1\over l^2}$:
\begin{equation}
\Xi (\kappa )={1\over 4\sqrt{\pi}}\int^\infty_{0} {ds\over s^{3/2}} 
\left( k_H (s)\Theta_S ({s\over \kappa} )-
\sqrt{\pi \over s}(({1\over 3}-{s\over 20})\kappa+{s\over 9})
\right)
e^{-{s/ 4}}~~.
\label{25}
\end{equation}
The constant $G$ in (\ref{24}) is the renormalized Newton constant.
For simplicity we  set to zero the renormalized 
values of the $R^2$ -couplings in the gravitational effective action
when derived (\ref{24}).
The expression (\ref{24}) is the generic expression
for the quantum corrected entropy of an
extremal black hole.  It may be understood as the
universal limiting function of the entropy of a non-extremal black hole (\ref{3})
in the extremal limit.
In particular, when $l=l_1$ the last term (given by (\ref{25}))
is an irrelevant constant and the entropy takes the form (\ref{1}) 
modified by a logarithmic term
\begin{equation}
S_{RN}^{ext}={A_+\over 4G}-{1\over 180} \ln {A_+\over \mu}~~,
\label{26}
\end{equation}
where $A_+=4\pi l^2$ is the horizon area.

A short digression is worth making here.  
There is a simpler way to get the expression (\ref{26}).
One should just notice that when geometry is characterized by just one dimensional 
parameter $l$ the latter can appear in the expression for the effective action
$W_{eff}$ (\ref{10}) and/or quantum entropy $S_q$ only in the ratio
${\epsilon /l}$ with the UV cut-off $\epsilon$. Thus all the dependence of
the entropy or effective action on $l$ can be extracted by knowing only the 
divergent part of $S_q$ or $W_{eff}$ respectively.
How it works for the extreme RN black hole is seen by comparing
 the eq.(\ref{2}) and eqs. (\ref{24'}) and (\ref{26}). This principle is universal and works
not only for extreme holes but for any geometry characterized by just a single
dimensional parameter. For extreme black holes one can exploit this to obtain
the entropy without even knowing  the limiting geometry.
This  observation is useful because (as we have already mentioned)
not all the known black holes fall in the class described by 
the metric (\ref{3}). In other cases the limiting procedure 
can be analyzed in the same way (see for example ref. \cite{91/2}). 
However for the entropy calculation, we just need to  know the 
UV divergence of the quantum entropy.

As an example consider the extreme dilaton black hole 
characterized by mass $m$ and electric charge $q$  
where the extremality condition implies $2m^2=q^2$.
The horizon area vanishes in this case. Therefore the 
classical black hole entropy is zero.
Nevertheless, the quantum entropy of the extreme dilaton hole 
does not vanish and has the UV divergent part calculated in \cite{2} 
(equation (29) in \cite{2}).
The above arguments suggest that the total renormalized entropy
of the extreme dilaton black hole is (up to an irrelevant constant)
\begin{equation}
S^{ext}_{dil}={1\over 18}\ln {q\over \mu}~~.
\label{27}
\end{equation}
Another example is provided by the extreme Kerr-Newman (KN) hole characterized
by zero electric charge $q$ and mass $m$ and
rotation $a$ related as $m^2=a^2$. Then, using the eq.(4.12) of ref.\cite{10}
we find  the total quantum entropy in this case to be
\begin{equation}
S^{ext}_{KN}={A_+\over 4G}+{1\over 90} \ln {A_+\over \mu}~~,
\label{28}
\end{equation}
where $A_+=4\pi m^2$.

In a similar manner we can extract some information about 
entropy even in non-extremal cases.
Using the expression for the divergence of the entropy of a RN hole
(equation (20) in \cite{2}) we find that the total quantum entropy
in this case after UV renormalization takes the form
\begin{equation}
S_{RN}={A_+\over 4G}+({2r_+-3r_-\over 90r_+})\ln {r_+\over \mu}+
\Upsilon ({r_-\over r_+})~~,
\label{29}
\end{equation}
where $r_+$ and $r_-$ are the outer and inner horizons.
The function $\Upsilon ({r_-\over r_+})$ can not be determined within 
this method.

\bigskip

Consider now how our results are modified to
include the class of black holes with non-trivial horizon topology.
The limiting geometry now is $E_g=H_2\times \Sigma_g$,
where $\Sigma_g$ is a compact constant curvature surface of genus $g$.
Calculating the effective action on $E^\alpha_g=H_2^\alpha\times \Sigma_g$
for arbitrary genus $g$ we need
just to replace in (\ref{15}) $Tr K_{S_2}$ by
the trace $TrK_{\Sigma_g}=\Theta_{\Sigma_g}$.
For arbitrary $g$ the formula for the quantum entropy then becomes
\begin{equation}
S^{ext}_{q, g}={1\over 4\sqrt{\pi}}\int^\infty_{\epsilon^2 / l^2} {ds\over s^{3/2}} 
k_H (s)\Theta_{\Sigma_g} (s {l^2\over l_1^2})e^{-{s/ 4}}~~
\label{31}
\end{equation}
generalizing the expression (\ref{21}) for the spherical horizon topology.
Remarkably, the heat kernel $K_{\Sigma_g}$ and its trace are  known
for arbitrary $g$.

The torus $T$ ($g=1$) with metric (\ref{x}) is the plane
${\cal R}^2$ factored by an integral lattice $\Gamma$, $T\simeq {\cal R}^2/\Gamma$,
that is, the points ${\bf z}$ and ${\bf z}+2\pi {\bf n}$, 
with $2\pi {\bf n} \in \Gamma$,
are identified.
The heat kernel on $T$ can be obtained from the heat kernel on 2d plane by taking the 
infinite sum over images. The trace of the heat kernel then takes
the form \cite{7}
\begin{equation}
\Theta_T(s)={A(T)\over 4\pi s}\sum_{2\pi {\bf n}\in 
\Gamma}e^{-{(2\pi {\bf n})^2\over 4s}}~~,
\label{30}
\end{equation}
where $A(T)$ is area of $T$.
The function $\Theta_T(s)$ is actually a multidimensional Jacobi theta function.
In the limit  $s\rightarrow 0$ it behaves as
$$
\Theta_T(s)={A(T)\over 4\pi s}+O(e^{-{1\over s}})~~.
$$
We may arrange the parameters of the lattice $\Gamma$ to have $A(T)=4\pi$. The 
area of the horizon then is $A_+=4\pi l_1^2$. The UV divergence of the 
quantum entropy (\ref{31}) for the toroidal horizon topology is
\begin{equation}
S_{div}^{g=1}={l_1^2\over 12\epsilon^2}-{1\over 15}{l_1^2\over l^2}\ln {l\over \epsilon}
\label{32}
\end{equation}
and the renormalized total entropy in the extremal limit reads
\begin{equation}
S^{ext}_{g=1} (l,l_1)={A_+\over 4G}-{1\over 15}
{l_1^2\over l^2}\ln {l\over \mu}+\Xi_{g=1} ({l_1\over l})~~,
\label{33}
\end{equation}
where the last term is function of only the ratio $\kappa={l^2_1\over l^2}$:
\begin{equation}
\Xi_{g=1} (\kappa  )={1\over 4\sqrt{\pi}}\int^\infty_{0} {ds\over s^{3/2}} 
\left( k_H (s)\Theta_T ({s\over \kappa} )-
\sqrt{\pi \over s}(({1\over 3}-{s\over 20})\kappa )
\right)
e^{-{s/ 4}}~~.
\label{34}
\end{equation}

The compact surface $\Sigma_g$ with genus $g\geq 2$ is 
identified with $H_2 /\Gamma$,
where $H_2$ is the hyperbolic space (usually represented 
as the upper half plane),
$\Gamma$ is a discrete subgroup of $SL(2, {\cal R})/ \{ \pm 1 \}$. 
The heat kernel on
$\Sigma_g$ then is determined from the heat kernel 
on $H_2$ (\ref{13}) by means
 the generalized procedure of summing over images.  
The trace of the heat kernel
can be calculated using
Selberg's trace formula \cite{11}, \cite{12}:
\begin{equation}
\Theta_{\Sigma_g}(s)=A(\Sigma_g ){e^{-{s/ 4}}\over (4\pi s)^{3/2}}
\int_0^\infty {dy y e^{-{y^2\over 4s}}\over \sinh{y\over 2}}+
{1\over 2}\sum_{n=1}^{\infty}\sum_{\gamma~ prim}{l(\gamma )\over
\sinh {n l(\gamma )\over 2}}{e^{-{s/ 4}}\over (4\pi s)^{3/2}}
e^{-n^2l^2(\gamma )/s}~~.
\label{35}
\end{equation}
Here the sum over $\gamma$ primitive indicates summing over all 
$\gamma \in \Gamma$
which are not powers of another element in $\Gamma$ with exponent $\geq 2$
(if $\gamma$ is primitive, $\gamma^{-1}$ is also counted as primitive),
and for each $\gamma$ the corresponding length of a closed geodesic
$l(\gamma )$ is given by $\cosh {l\over 2}=|tr \gamma |/2$. For  area
of $\Sigma_g$ we have  $A(\Sigma_g)=
2\pi |\chi (\Sigma_g )|=4\pi (g-1)$.
Only the first term in (\ref{35}) has a pole for small $s$
so that
$$
\Theta_{\Sigma_g}={(g-1)\over s}(1-{s\over 3}+O(s^2))~~.
$$

This helps us to find   the UV divergence of the quantum entropy (\ref{31})
\begin{equation}
S_{div}^{g\geq 2}={l_1^2\over 12\epsilon^2}-(g-1)({1\over 18}+
{1\over 15}{l_1^2\over l^2})\ln {l\over \epsilon}~~.
\label{36}
\end{equation}
After  renormalization the total entropy of a black hole with higher genus
horizon topology takes the following form in the extremal limit 
\begin{equation}
S^{ext}_{g\geq 2} (l,l_1)={A_+\over 4G}-(g-1)({1\over 18}+{1\over 15}
{l_1^2\over l^2})\ln {l\over \mu}+\Xi_{g\geq 2} ({l_1\over l})~~,
\label{37}
\end{equation}
where 
\begin{equation}
\Xi_{g\geq 2} (\kappa )={1\over 4\sqrt{\pi}}\int^\infty_{0} {ds\over s^{3/2}} 
\left( k_H (s)\Theta_{\Sigma_g} ({s\over \kappa} )-
(g-1)\sqrt{\pi \over s}(({1\over 3}-{s\over 20})\kappa-{s\over 9})
\right)
e^{-{s/ 4}}~~.
\label{38}
\end{equation}

The expression (\ref{21}) and its generalization (\ref{31})
are our main results. They give the universal behavior of the quantum entropy
of a black hole in the extremal limit. The UV divergences of the entropy (\ref{23}),
(\ref{32}) and (\ref{35}) are in agreement with the general formula given in 
\cite{2}. The equations (\ref{21}), (\ref{31}) can be generalized to the metric
(\ref{8}), (\ref{xxx}) for arbitrary values of the parameter $c$.
This, however, involves problem of the calculating the heat kernel
on the 2d sub-space $(x,\phi )$ which is not a space of constant 
curvature for $c\neq 1$. It is interesting to note that the entropy (\ref{21}),
(\ref{31}) is completely determined by the spectral geometry of the horizon
surface $\Sigma_g$ that is encoded in the theta function 
$\Theta_{\Sigma_g}(s)$. In principle, we could repeat our calculation 
for a black hole whose horizon is an arbitrary not necessarily symmetric
2d surface of genus $g$. Our general expression (\ref{31}) for the entropy
is valid for this case as well. However, the calculation of the function
$\Theta_{\Sigma_g}(s)$ becomes considerably more complicated. We may conclude 
that in the extreme limit the quantum entropy of a black hole is
effectively ``two-dimensional'' and is determined by the 
geometry of the horizon. This property of the entropy becomes even more apparent in the 
case of so-called ultra-extreme black holes for which $b=0$ in (\ref{4}) in the
extreme state \cite{13}.
The extreme geometry \cite{5}
\begin{equation}
ds^2=x^2d\phi^2+dx^2+r^2_+d\omega^2_{(g)}
\label{40}
\end{equation}
then describes the direct product space $U={\cal R}^2\times \Sigma_g$ of 2d plane
${\cal R}^2$ and compact surface $\Sigma_g$. The whole quantum analysis we 
have done before is simplified for the metric (\ref{40}).
This is because the heat kernel on the conical plane ${\cal R}^2_\alpha$
has the simple form
\begin{equation}
Tr K_{{\cal R}^2_\alpha}=\alpha Tr K_{{\cal R}^2}+d(\alpha )~,~~
d(\alpha )={1-\alpha^2\over 12\alpha }~~.
\label{41}
\end{equation}
It can be obtained by taking accurately the limit $l\rightarrow \infty$
in the equation (\ref{19}).
Using this expression we find for the effective action
on $U^\alpha$
\begin{equation}
W_{eff}[U^\alpha ]={1\over 2}\ln \det (-\Box )|_{U^\alpha}=
\alpha \ln \det (-\Box )|_{U}+d(\alpha )\ln \det (-\Box )|_{\Sigma_g}~~,
\label{42}
\end{equation}
where the last term is the logarithm of determinant
of the Laplace operator $-\Box $ 
defined on $\Sigma_g$. The quantum entropy 
then takes an especially simple form
\begin{equation}
S^{uext}_q={1\over 12}\int^\infty_{\epsilon^2\over r^2_+}{ds\over s}
\Theta_{\Sigma_g}(s)
\label{43}
\end{equation}
and is expressed as the determinant of the two-dimensional Laplace operator
\begin{equation}
S^{uext}_q=-{1\over 6}\ln \det (-\Box )|_{\Sigma_g}
\label{44}
\end{equation}
considered on the horizon surface $\Sigma_g$.
So, as we can see from (\ref{44}), the quantum entropy is purely
``two-dimensional'' in the ultra-extreme case.

\bigskip

To summarize, we considered the extreme limit of a black hole geometry
of the RN type.
Universally, the limiting geometry is the direct product of two 2-dimensional
spaces and is characterized by just a few parameters corresponding to
their length scales and the amount by which one of them is not 
a space of constant negative curvature (the other 2-space has been assumed
to have constant curvature).   The quantum entropy
of a massless scalar field possesses a corresponding universality
in the extremal limit. The form of the limiting entropy is found (\ref{21})
as function of the parameters of the limiting geometry. We have generalized
it to  black holes whose horizons have toroidal or higher genus topology 
(\ref{31}). In general, the extreme quantum entropy 
is completely determined by the spectral geometry of the horizon. In the
ultra-extreme case it is just the determinant of the two-dimensional
Laplacian. As a byproduct of our consideration we obtained expressions for the
quantum entropy of black holes which are not of the RN type:
extreme dilaton and extreme KN holes. In both cases the classical BH entropy
is modified by some logarithmic corrections.
Our results do not apply directly
to extreme black holes in  superstring theory \cite{4}
though an appropriate extension to this case
might be an interesting problem.

\bigskip

{\it Note added:}

After this paper was complete we became aware of the recent preprint
\cite{21} where the quantum entropy is found to vanish in the
extremal limit. We comment here on why this apparent disagreement 
is not in contradiction with our results.

In ref. \cite{21} the quantum thermodynamical entropy
is calculated by making use of the equation (eq.(10) in \cite{21})
$\partial_{r_+}S_q= \eta$ where
$\eta$ is constructed from the renormalized stress-energy tensor.
Integrating this equation 
$$
S_q=\int^{r_+}\eta + C
$$
produces an arbitrary constant $C$. This is a reasonable
ambiguity since the thermodynamical entropy is always defined up to 
an irrelevant constant. However in ref. \cite{21} this ambiguity is used 
to choose $C=C(r_B)$ so that for a generic non-extreme hole
the entropy (eq.(11) in \cite{21}) is defined to be 
$$
S_q=\int^{r_+}_{r_B}\eta 
$$
which vanishes when the size of the system $(r_B-r_+)$ tends to zero. 

The system under consideration represents a rather complicated interaction of 
two objects: a black hole and a hot gas (for a discussion of this see ref.
\cite{51/2}). Both objects contribute to the
thermodynamical entropy $S_q$. The contribution of the hot gas can be
identified and eliminated through its dependence on the size of the system.
On the other hand, the contribution of the hole itself does
not depend on this size. The method for calculating the entropy 
that we employ gives this contribution automatically.
We believe that the vanishing of the entropy in the extremal case 
found in ref. \cite{21}
is a result of the above definition of $S_q$ that was used there.
First of all, it contains both the black hole and hot gas
contributions. Moreover, by adding the constant $C(r_B)$ 
(irrelevant for the entropy of a non-extreme hole)
one effectively subtracts the contribution of the hole itself
from the entropy of the extreme hole ($r_B=r_+$) in the definition of 
ref. \cite{21}.  By keeping $C$ an arbitrary
constant and extracting the $r_B$-independent part  
in $S_q$ one would obtain the correct expression 
for the entropy of the hole itself that does not vanish,
reproducing  our result in the extreme limit. 

\section*{Acknowledgements}

This work was supported by the Natural Sciences and 
Engineering Research Council
of Canada and by a NATO Science Fellowship.

\end{document}